# Conditions for entangled photon emission from (111)B site-controlled Pyramidal quantum dots

G. Juska[*], E. Murray, V. Dimastrodonato, T. H. Chung, S. Moroni, A. Gocalinska and E. Pelucchi

*Tyndall National Institute, University College Cork, Lee Maltings, Cork, Ireland*

**Abstract**

A study of highly symmetric site-controlled Pyramidal $In_{0.25}Ga_{0.75}As$ quantum dots (QDs) is presented. It is discussed that polarization-entangled photons can be also obtained from Pyramidal QDs of different designs from the one already reported in Juska et al. (Nat. Phot. 7, 527, 2013). Moreover, some of the limitations for a higher density of entangled photon emitters are addressed. Among these issues are (1) a remaining small fine-structure splitting and (2) an effective QD charging under non-resonant excitation conditions, which strongly reduce the number of useful biexciton-exciton recombination events. A possible solution of the charging problem is investigated exploiting a dual-wavelength excitation technique, which allows a gradual QD charge tuning from strongly negative to positive and, eventually, efficient detection of entangled photons from QDs, which would be otherwise ineffective under a single-wavelength (non-resonant) excitation.

---

[*] E-mail: gediminas.juska@tyndall.ie

I.   **INTRODUCTION**

A number of possible routes are currently under investigation with the aim to find a practical, technological implementation of quantum processing. One of the promising options is that of building a quantum processor based on photonic technologies. Nevertheless, options and alternatives branch readily even when the photon source problem endures, and there is no general agreement in the community on what alternatives will finally win this race.

In the field of entangled photon sources, two main options exist at the moment. One of them relies on non-linear optical processes[1], with the advantage of being generally highly efficient in photon entanglement (and photon throughput) while allowing operation at room temperature. However, the technology does not deterministically guarantee a single pair of entangled photons on-demand. Moreover, it is reasonably difficult to integrate in a photonic chip with current non-linear crystals or equivalent implementations, as it remains a relatively bulky technology, however there are relevant results in this direction.[2,3] On the other hand, quantum dots (QDs) are compatible with semiconductor foundry technologies, allow true photon on-demand operation, but operate at cryogenic temperatures and have not shown, untill now, the same, reproducible, high entanglement quality as non-linear sources[4,5,6,7,8,9]. Some important milestones have been met[10,11] and the progress is proceeding lively. Nevertheless, despite the significant advancement of the QD technologies, some important issues need to be addressed: e.g. a truly integrable (and scalable) system should enable site-control at the epitaxial stage, as it would allow pre-aligning the QD with a semiconductor photonic circuit. Site-control is a necessary feature, as the photonic circuit architecture is likely to be composed of billions of gates.

As proposed originally in Benson's et al. manuscript, in analogy to an atomic system singlet state recombination[12,13], the entangled photon emission from QDs relies on the formation of an entangled atomic state between two identical particles (excitons in this case) occupying two nearly degenerate levels (and forming a biexciton).

We have recently shown[14] that matching site-control and entanglement preservation is indeed a possibility with specially grown Pyramidal QDs. They are grown on (111)B patterned substrates by metalorganic vapour phase epitaxy (MOVPE) due to anisotropies in the metalorganic precursors decomposition process and what have been reported as capillarity effects[15,16,17]. The intrinsic lattice symmetry[18,19] associated with the growth direction is the basis of the creation of highly symmetric dots, as pointed out in a number of reports.[20,21] Yet, until our report[14], no group could actually obtain entangled photons from (111) site-controlled dots, questioning the very possibility of attaining such result. Our recent results were obtained only due to a specific growth procedure, exposing the QD to unsymmetrical dimethylhydrazine (U-DMHy) (a nitrogen precursor) allowing a capillarity-induced, formation process.[22] The surfactant effect[23] of U-DMHy allowed, to get relatively high density of symmetric dots, with regions where around 15% of the orderly positioned dots emitted entangled photons.[14]

Despite the striking impact of this outcome on the technological improvement for quantum processing, it would be important to achieve such a result without necessarily relying upon a "special" trick, and in a much general manner. Moreover, an improvement of the basic understanding on the growth process would clarify the unexpected sources of asymmetry, and make entangled photon emission from the full family of QDs on (111) substrates a feasible achievement. This would hopefully allow tuning emission energies in a broad range of wavelengths and obtaining tailored emission properties, and not "constrained" ones.

We show here that it is also possible to obtain entangled photon emission from pyramidal dots (even if with a lower density of good emitters) without the exploitation of surfactant effects. We experimentally analyse some of the factors that limit the density of entangled photon emitters and their impact on our system. One of them is the usually small deviation of carrier confinement potential symmetry from the theoretically predicted three fold rotational one. The consequence of this deviation is a small exciton level splitting, also known as a fine-structure splitting (FSS): a usual issue of the most QD systems, however, far less significant in Pyramidal QDs. Another, more important limiting phenomenon occurs when the system is non-resonantly excited – Pyramidal QDs tend to charge efficiently with negative charge carriers from surrounding material. This, as will be discussed, dramatically reduces the amount of useful biexciton-exciton recombination cascade events. This issue is addressed in the paper and a possible solution, achieved by a dual-wavelength excitation, is presented.

Our manuscript is organised as follows. We start with a brief background on the physics of entangled photon emission from QDs (providing the general reader with a proper background on the current status). After the experimental summary, we discuss how it is possible to obtain photon entanglement by a variety of epitaxial growth recipes, and discuss the limits for the performances in samples which were grown without surfactants effects. We explain one of the the underlying phenomenology of the found performance limits, negative charging of QDs, which prevents entangled photon emission cascades. Finally, we show a practical solution to the problem: dual-wavelength excitation.

## II. Theoretical background notes

The original proposal to obtain entangled photon emission from QDs[13] is based on an analogy to atomic systems, where, in the early 1970s, entangled photon emission was demonstrated, for example, by exploiting an atomic cascade from a specific atomic singlet (in a p state) in a calcium atom[12]. Shortly, in atoms, in a singlet state, entanglement of the electronic wavefunction is the result of textbook particle indistinguishability, which forces the atomic system description to be antisymmetric (two fermions) in respect to particle exchange. Symbolically, without normalization, $|\psi_{atom}\rangle \propto (|\uparrow,\downarrow\rangle - |\downarrow,\uparrow\rangle)$, where the arrows conventionally indicate the spin degree of freedom. The subsequent electronic jumps to ground state produce entangled photon emission, as the two photon emission simply directly maps the entangled electronic state. Should the electronic states occupied by the two electrons be non-degenerate, the entanglement would persist, i.e. the two photons would be constantly entangled over time, but in an entangled state that evolves in time due to the time-dependent phase induced by the non-degeneracy. This has been observed historically, for example, in what have been referred to as quantum beats (see for example Ref. 24), namely the energy differences between the atomic levels lead to a different time evolution of the (two) single electronic states, resulting in phase terms appearing in the entanglement probability amplitudes after emission (see also for example the recent Ref. 25). The physics formalism involved is in its essence similar, for example, to the beating in the ammonia molecule often treated in introductory textbooks to quantum physics (the vibrational "inversion" states for the first maser), and has strong similarities to Rabi oscillations.

In the case of QDs the picture is slightly more complicated. The decaying state would be a biexciton, i.e. a QD filled by 4 particles (fermions), two electrons and two holes, confined by the barriers and interacting through coulomb interactions. This does not impede particle indistinguishability and the need for a correct state description of appropriate parity under particle exchange in order to fulfil Fermi statistics (separately for the two electrons and the two holes). Obviously the description could also be that of a "singlet" state of two excitons, instead that of a state of four fermions, and both equivalent descriptions can be found in the literature.

The fundamentals of entanglement (particle indistinguishability and symmetry exchange requirements) cannot obviously be lifted, and the non separability of the biexciton state is not questionable (we indeed realised that, in the broad semiconductor community, the idea that the biexciton is separable, i.e. can be written as a direct product of the single particle states, is often appearing: it comes without saying that this should only be considered as a practical approximation if appropriate, and not as a complete physical description). We refer the interested reader to the extensive literature in the field, giving some examples in references 26, 27, 28, 29, 30, 31, 32, 33.

As a result, the biexciton in a QD behaves like in an atom, an artificial atom in this case: the biexciton photon cascade, through the electron-hole recombinations, produces entangled photons by merely maintaining/mapping the entanglement nature of the original (singlet like) electronic state. If the dot is perfectly symmetric the excitonic states are degenerate in energy and the emitted state is $|\psi\rangle = \frac{1}{\sqrt{2}}(|RL\rangle + |LR\rangle)$, with L and R indicating left and right circular polarization. If the dot is not symmetric the level degeneracy is lifted and "beating" appears, as in the atomic case. The final photon state will be $|\psi\rangle = \frac{1}{\sqrt{2}}(|HH\rangle + e^{i2\pi FSS\tau/h}|VV\rangle)$, with H and V standing for horizontal and vertical linear polarization, FSS is the difference between the two excitonic energy states we discussed in our introduction, and τ is the time between the first and the second exciton emission. This was pointed out in QDs, for the first time, in Ref. 34.

For QD's since the phase term is dependent on the emission time (typically in the nanosecond region), which is randomly distributed, the experimental state identification becomes complicated. When a FSS splitting is present the state tomography procedure[35] averages over several randomly distributed/emitted different entangled states, practically resulting in an apparent classical state. Effectively, only very small FSS (less than a few μeV) allow detection of entanglement without the need for post selective time resolved/windowing measurements.

We would like to caution the general reader on a specific terminology aspect. In the early history of QD entanglement development it could not be demonstrated because the FSS was too substantial. This has somehow generated a distinct jargon in some authors: since the asymmetry of the dot breaks the degeneracy of an intermediate exciton level, this potentially enables the two paths to be distinguished by frequency[36], i.e. a "which path" information is introduced which pre-empts entanglement. While this has been used as a jargon by specialised researchers with a specific contextual meaning we caution that this can be misleading. As it is clear from our discussion, entanglement is preserved in each single realization of the experiment (i.e. in each single cascaded emission). The non-degeneracy introduces a specific time-evolution of the two-photon state, which however remains strictly non-separable at all times. The non-degeneracy simply hinders the detection of such entanglement only in a statistical sense. Indeed, as already hinted above, to detect entanglement one needs to perform a full quantum tomography of the density matrix of the two photons, which implies several repetitions of the cascaded emission, both because several correlations must be measured and in order to have sufficient signal-to-noise ratio. If however the time-dependent phase induced by the non-degeneracy varies randomly from repetition to repetition, the result will be that of averaging out all phase terms in the density matrix and one will be left with a statistical mixture (i.e. classical correlations, non-entanglement). Rigorously, and outside specialised scientific jargon, in each specific repetition of the experiment there is no "which path" information in the cascade process, as only *after* the first photon is measured the superposition entangled state is projected onto a specific polarization and energetic state. During the cascade and the "flying" period, it stays as an entangled state. The process has, for this reason, no real similarity with a "double slit" experiment where the slit the photon has gone through is known (or the alike process in a Mach-Zehnder type set-up and equivalents) where a "which path" information obtained by some extra/external measurement can be effectively collected. It would, on the other hand, have resemblances, if any, with the phenomenon of coherence loss caused by random phases[37].

## III. Experimental techniques

The results presented in this work are obtained from a batch of eight $In_{0.25}Ga_{0.75}As$ samples grown without and two samples with surfactant effects and of different epitaxial QD design, namely QD thickness and growth temperature, as presented in Table 1. First, all the samples were characterised by measuring the fine-structure splitting dependence on a QD thickness (QD emission energy) and the surfactant effects. Second, five samples with the smallest FSS values were selected and shown as sources of entangled photon emitters. It should be said that finding good emitters did not require a particularly extended amount of time (given the difficulty of the task in "normal" conditions), and a few good dots per sample could be found during a one day search. In these samples the density of good dots does not match what we reported in Ref. 14, an indication that more work is needed to realize extended arrays of emitters. We underline that all the samples are representative of different growth conditions for the QD/barriers structure only: the alloy composition of cladding ($Al_{0.55}Ga_{0.45}As$)/confining (GaAs) and QD ($In_{0.25}Ga_{0.75}As$) layer was kept the same (we refer the reader to Ref. 38 for more details). The dot thickness and/or dot/barriers growth temperature were changed as reported in Table 1. For temperature values different than the reference sample (730 °C nominal), slow ramping steps were performed during the deposition of the cladding layers such to start the epitaxial growth of the barriers (and therefore of the QD) with a stable temperature condition. Growing the barrier/QD structure at a constant temperature is an important aspect for pyramidal QDs as different temperatures will deliver different dot shape/size, due to the different equilibrium between diffusion-induced capillarity and growth rate anisotropy, as we have recently reported[15]. For example the dot base dimension will change from ~30 nm to 50 nm, when the temperature is changed by 60 degrees, resulting in significantly different confinement and, in general, dot properties.

**TABLE 1.** Growth conditions and parameters for the discussed pyramidal $In_{0.25}Ga_{0.75}As$ QD samples. The temperature values are the thermocouple readings.

|        | #1  | #2  | #3  | #4   | #5  | #6 | #7  | #8   | #9   | #10 |
|--------|-----|-----|-----|------|-----|----|-----|------|------|-----|
| h [nm] | 0.5 | 0.5 | 0.5 | 0.57 | 0.8 | 1  | 1.5 | 1.75 | 0.85 | 2   |
| T [°C] | 640 | 700 | 730 | 730  | 730 | 730| 730 | 730  | 730  | 730 |
| U-DMHy | no  | no  | no  | no   | no  | no | no  | no   | yes  | yes |

All samples were measured in apex-up geometry, which requires a substrate removal procedure[38,39]. Photoluminescence data were taken in the conventional micro-photoluminescence set-up, which enabled access to individual QDs. The samples were cooled down to 8 K by a closed-cycle helium cryostat. QDs were excited non-resonantly with a semiconductor laser diode emitting at 635 nm. Exciton and biexciton transitions were filtered for correlation measurements by two monochromators. Each filtered transition was divided by a polarizing beamsplitter and sent to silicon avalanche photodiodes (APD). Four synchronized sequences of APD signals were fed to photon counting module and analysed to build four the second-order correlation curves $g^{(2)}(\tau)$. Excitation wavelength dependent studies were carried out using a supercontinuum fiber laser equipped with an acusto-optical filter which enabled a simultaneous selection of up to eight different laser emission wavelength values in the range between 600 and 1100 nm.

To describe a two-photon (namely a biexciton and an exciton) polarization state, we used a quantum state tomography procedure[35] to measure the density matrix $\rho$. As the expected maximally entangled state of photons emitted from a QD is $|\psi\rangle = \frac{1}{\sqrt{2}}(|H_{XX}H_X\rangle + |V_{XX}V_X\rangle)$, the fidelity $F = \langle\psi|\rho|\psi\rangle$ to this state can be calculated without performing a full tomography procedure – only five out of sixteen two-photon Stokes parameters are required and they can be obtained from the correlation measurements in linear (L), diagonal (D) and circular bases (C)[40]. If there is no in-plane polarization anisotropy, the fidelity can be expressed by only three two-photon Stokes parameters – degrees of correlations ($C_{basis}$): $F = (1 + C_L + C_D - C_C)/4$, where $C_{basis} = (g^{(2)}_{xx,x} - g^{(2)}_{xx,\bar{x}})/(g^{(2)}_{xx,x} + g^{(2)}_{xx,\bar{x}})$.[41] For simplicity, in most of the cases we used the fidelity marker (> 0.5 for non-classical light) to show polarization-entanglement.

## IV. EXPERIMENTAL RESULTS AND DISCUSSION

### A. Entangled photon emitters from quantum dots of different epitaxial designs

As we discussed, in the contest of a search of improved development of site-controlled QD sources of entangled photons, it is important as the first step to investigate the conditions over which a certain epitaxial recipe is capable of delivering symmetric dots.

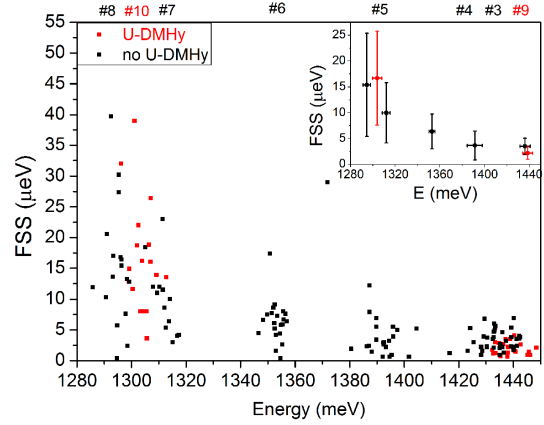

**FIG. 1.** The fine-structure splitting as a function of exciton transition energy of the samples grown at 730°C. The numbers at the top of the graph indicate the samples presented in Table 1, each contining a high number of site-controlled dots. (Inset) Average FSS values for the dots in the samples. Sample #3 was omitted from the statistical calculations due to a small number (10) of measured QDs. Error bars are standard deviations.

In FIG.1 we show the fine-structure splitting values measured from the samples grown at 730°C. The values are plotted as a function of an exciton emission energy which reflects the real QD thickness. We avoid using nominal values, as QDs exposed to U-DMHy emit at higher energy than the regular counterpart QDs of the same nominal thickness. In this case, we assume that U-DMHy acts as a surfactant and one of its effects is a small reduction of QD thickness and thus increased confinement effects. The measured FSS dependence on the emission energy shows that the FSS values strongly depend on the QD emission energy. FSS and the spread of its values non-monotonously increase as the QDs get thicker – from 3.5±1.6 μeV to 15.4±10.0 μeV for 0.5 nm and 1.75 nm thickness of regular QDs, respectively (the inset of FIG. 1). While vanishing FSS of QDs grown along (111) direction is predicted theoretically[20,42,43] and typically small values were obtained experimentally[21,44,45], there are a number of effects that can cause deviation from theory. QD alloy disorder is a potential cause of reduced symmetry[46]. Moreover, QD environment is an important factor, especially as pyramidal QDs are a part of a complex interconnecting ensemble of nanostructures, and the confinement potential profile is non-trivial. For example, we find that QDs similar to the ones from sample #3 but confined by $Al_{0.3}Ga_{0.7}As$, have FSS of 58.7±25.4 μeV compared to 3.5±1.6 μeV of their counterparts confined by GaAs. However, we stress that different barrier material not only changes the confinement potential height and profile, but it has strong effect on QD size, aspect ratio[47], thus the change of FSS cannot be solely attributed to the barrier material. For comparison reasons, the FSS distribution of samples grown by exposing a QD layer to U-DMHy (#9 and #10) is shown in FIG. 1, as well.

In FIG. 2(a) we show representative spectra of five samples, where QD entangled photon emitters were found. The fidelity values of the expected maximally-entangled state are presented next to the corresponding spectra, with the highest measured value of 0.622±0.017 in pulsed excitation mode. Two main parameters were varied. The nominal QD layer thickness was varied for samples #3, #4 and #5, which were grown with the same conditions concerning anything else in the structures. On the other hand, samples #1, #2 and #3 were grown with the same dot thickness, but different QD nominal temperatures, namely 640, 700 and 730 ˚C, keeping again all the other structural parameters the same.

We stress that growth temperature affects the width of the self-limiting profile which is the base for the QD formation[16,17]. In this case, the self-limiting profile width (otherwise the QD base) of the lower GaAs confining barrier is changing from ~30 to ~70 nm (the temperature range is from 640 to 730 ˚C, respectively). As discussed in Ref. 47, the overall effect is reduction of the QD volume when the sample is grown at lower temperatures. In this particular case, the QD emission energy could be tuned in an average range of 40 meV.

FIG. 2 (b, c) presents polarization resolved second order correlation curves taken in linear, diagonal and circular bases from representative dots of the samples #5 and #2. The selection of the bases is a conventional way to demonstrate polarization entanglement based on the fact that the two-photon polarization state $|\psi\rangle = \frac{1}{\sqrt{2}}(|HH\rangle + |VV\rangle)$ can be expressed in diagonal and circular bases as $|\psi\rangle = \frac{1}{\sqrt{2}}(|DD\rangle + |AA\rangle) = \frac{1}{\sqrt{2}}(|RL\rangle + |LR\rangle)$. In the presented cases, biexciton and exciton polarization state correlations were observed in linear and diagonal bases, while the anti-correlation of the two photon states emitted in the recombination cascade was detected in circular basis. The fidelity of the expected maximally entangled state of the photons emitted from one of the QDs of sample #5 was calculated to be 0.622±0.017. This particular dot had a FSS of 1.3±0.5 μeV. After applying a simple time-gating technique[34] (not shown), the fidelity value increased to 0.738±0.020 at the price of reduced two-photon intensity by ~20% (the gate width was 3ns). The curves from the samples #1 and #2 were taken in a continuous-wave excitation mode.

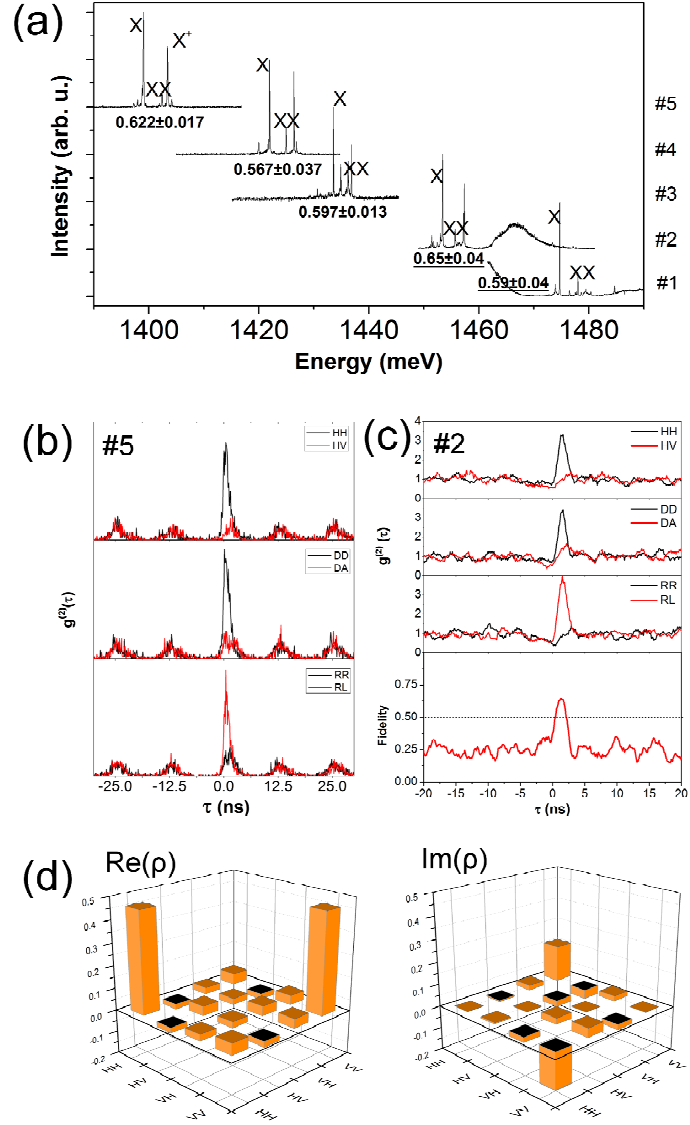

**FIG. 2.** (a) Representative spectra of five different design samples with QDs emitting polarization-entangled photons. Entangled state fidelity values are next to the corresponding spectrum. (b) Polarization-resolved second-order correlation curves taken in linear, diagonal and circular bases; sample #5. (c) Polarization-resolved second-order correlation curves taken under continuous wave excitation; sample #2. (d) The real and imaginary parts of the two-photon polarization state density matrix obtained from a QD with a FSS of 2.9±0.2 μeV from sample #5.

The potential for entangled photon emitters can be discussed showing the distribution of FSS values (FIG. 2). The fidelity to the expected maximally entangled state measurements showed that the bottom limit of a non-classical light source (0.5) typically was obtained with QDs with FSS equal to ~2-3 μeV. We stress that fundamentally this limit can be bigger, as shown in FIG. 2 (d), where a full density matrix of the two-photon polarization state of a QD with a FSS of 2.9±0.2 μeV is presented (sample #5). Because of the FSS, a maximally entangled state tends to be different (in this case $|\psi\rangle = \frac{1}{\sqrt{2}}\left(|HH\rangle + e^{0.41\pi i}|VV\rangle\right)$) from the expected one, as it is proven by non-vanishing off-diagonal elements in the imaginary part.

The only sample, in the discussed ensemble, which has an average FSS (2.1±1.2 μeV) very close to the FSS limit to prove entanglement without time-gating technique is #9, grown with U-DMHy, and previously reported in Ref. 14. The next smallest value (3.5±1.6 μeV) is of that of sample #3, which emits at nearly the same energy as #9. While the FSS difference is very modest, experimentally we interestingly observed far smaller density of entangled photon sources from sample #3 compared to sample #9. While exposure of thicker QD layers to U-DMHy does not have a major positive impact (sample #10), we argue that for thin QDs it can play a significant role. The surfactant effects can reduce the FSS by a few μeV – an amount sufficient to reach the limit when polarization-entanglement can be observed without external FSS tuning or time-gating procedures. On one hand, this observation confirms that the surfactant effects caused by the exposure to unsymmetrical dimethylhydrazine are favourable. On the other hand, it is clear that the effects are not essential in order to observe entangled photon emission from Pyramidal QDs and a simplified growth procedure can be used. In either case, in the future, a local FSS tuning strategy, such as an applied strain and/or electric field[48], will be required to increase the density of bright sources of entangled photon emitters.

## B. Excitonic pattern dispersion and characteristics

As discussed above, the fine-structure splitting, even though small, is one of the limiting factors of this QD system. In the further discussion, we would like to show an additional, but not fundamentally limiting, factor – negative charging of QDs – and a possible solution to it.

A mutual feature of all QDs found as entangled photon sources in this work, and the ones reported in Ref. 14, is the same charging configuration which reflected in the excitonic pattern. As we will discuss, it appears from our data that the population of these QDs is dominated by *positively* charged carriers. A QD spectrum featuring a positive trion transition was a reliable indicator of potential entangled photon emitters. The total density of such positively charged QDs in our any sample (irrespective of other variables such as QD alloy composition, thickness, growth temperature) varies from 25% to vanishing values. We anticipate that this pattern is not in itself a unique signature of highly symmetric dots, as obviously expected. As it will be discussed later, it is simply a consequence of a specific QD charging mechanism, which causes, on the other hand, the majority of non-resonantly excited QDs to be negatively charged.

In most of the cases the efficiency of *negative* charging is such that the two-photon intensity of the biexciton-exciton recombination cascade becomes practically useless for entangled photon emission, even though the QD itself is highly symmetric. Only the dots which had nearly balanced capture rate of electrons and holes, or dominant charging by holes, were practically suitable for entanglement measurements.

We emphasize that an unambiguous indication of the charging type could in principle be made by charging a QD integrated in a light-emitting diode type structure[49]. Unfortunately this is not available at this stage. It still needs developing in the Pyramidal QD system, as the non–planarity of the system complicates sample processing and design. In this work, the attribution of charging type is solely based on "equivalent" experimental observations and theoretical insights reported in literature. While in theory there are no limitations to very different combinations of the excitonic transitions energetic ordering, in practice, a positive trion has been usually reported/observed at a higher energy than the exciton (a review can be found in Ref. 50). Moreover, we found a good agreement with positive charge configurations, and specifically, the fine structure of a 'hot trion'[51,52,53], which we as well used to identify the observed transitions by photon-correlation measurements.

1. **Negatively charged QDs**

As discussed, non-resonantly excited QDs with a positive (or balanced) charging appeared to be of great importance for the pre-selection of potential emitters of entangled photons. Depending on the sample preparation and the QD design, when the pre-selection of positively charged dots was performed, the percentage of good emitters could be as high as 75% (sample #9), or sometimes lower, 10-15% as in sample #5. In itself a very high density of good dots if compared to any other QD system reported to date.

However, we would like to discuss that there is space for improvement. In fact, the overall density of good dots (when one includes all patterned dots in the count) was never higher than 25%, implying that there is a relatively big limitation in the system, as the remaining dots appeared to be practically useless for photon-correlation measurements. This issue is addressed in this section.

A colour map in FIG. 3 (a) presents an excitation power dependence of excitonic transitions and their intensity taken at low power for long (10 s) integration of a representative dot from sample #5. Similarly to the dots used for entanglement tests, the first transitions to appear are positively charged (X+) and neutral excitons (X). However, the positive trion is very quickly suppressed by the appearance of a new transition on a lower energy side of X. The spectrum at the stage when a biexciton (XX) appears is shown in FIG. 3 (c), as indicated by the arrows. At this excitation level the exciton saturates and only the intensity of X- and XX are increasing (the validity of the attribution of X and XX was confirmed by a well-pronounced bunching in the second-order correlation function, FIG. 3 (b), and by a measured fine-structure splitting of 11 μeV, FIG. 3 (d): the FSS is reflected in both the exciton and the biexciton peaks, but not in charged peaks). The spectrum of the specific QD, presented by a red curve in FIG. 3 (c), was taken at relatively high excitation power. Its intensity is at comparable levels to the ones typically used in entanglement tests with positively charged QDs. This implies that the overall photon-correlation measurement procedure is very inefficient, as the most favourable condition for a high visibility of the bunching in the second-order correlation function is a high exciton/biexciton intensity ratio, or otherwise a low excitation power mode, similar to the one in FIG. 3 (c) described by a black curve (that these are the most favourable conditions for our QDs is confirmed by our systematic observations, similar to what reported, both experimentally and theoretically, in Ref. 54). In this chosen representative case, the very low X and XX intensity would increase the entanglement measurement collection time by a significant factor, making it nearly impossible to perform practically.

We emphasize that this reported example is actually a rather favourable one, and it was selected here only because of the exciton intensity when pumped at low excitation power (i.e. with a still "dominant" exciton) which allowed carrying out cross-correlation and the FSS measurements to prove the type of transitions. Unfortunately, the exciton intensity in the majority of the other negatively charged QDs was weaker in most samples, making these dots practically useless under non-resonant excitation for entangled photon emission.

At this stage it seems that the observed exciton intensity suppression is the consequence of an efficient QD feeding by electrons from the barrier material for some of the dots. Fast capture of an additional electron causes a negative exciton to be dominant over the neutral one. Further increase of the excitation power causes, as expected, the appearance of a neutral biexciton (a similar combination of these two dominant transitions, namely trion and neutral biexciton, could be easily obtained, reportedly, from QDs placed in a diode structure[55]). Nevertheless the neutral biexciton decay seems to result in a neutral exciton, which is efficiently charged by an electron before actually decaying and emitting a photon, depleting the neutral exciton spectral signatures.

To complete our discussion, we stress that the no-observation of a negatively charged biexciton is not a trivial outcome. It probably can be explained either because of a reduced/increased electron/hole capture rate due to Coulomb interactions or because of the absence of excited states in the QD conduction band. In either case, it should be said that a negatively charged exciton somehow closely resembles in our spectra a neutral exciton which is emitted in a biexciton-exciton recombination cascade and, actually, rather well-pronounced bunchings were obtained in cross-correlation curves between XX and X-, making the correct identification a non-trivial task.

The origin of this charging which prevents proper entanglement detection is not obvious, as all heteroepitaxial layers are nominally semi-insulating. A possible QD population scenario proposed in Ref. 56 states that negative charging occurs when the structure is excited non-resonantly with photon energy capable to transfer electrons from GaAs acceptor levels to the conduction band. For example, MOVPE grown GaAs is known to have a residual carbon acceptor level[57] which is reflected as an intense photoluminescence feature at ~1493 meV. In this way, an excess concentration of free electrons that can charge QDs can be created during laser excitation. It should be also said that we observed that higher concentrations of negatively charged QDs tend to be found in the areas where the GaAs substrate is completely etched away from the pyramids. Processing induced defects could possibly act as hole trap states and intensify negative charging.

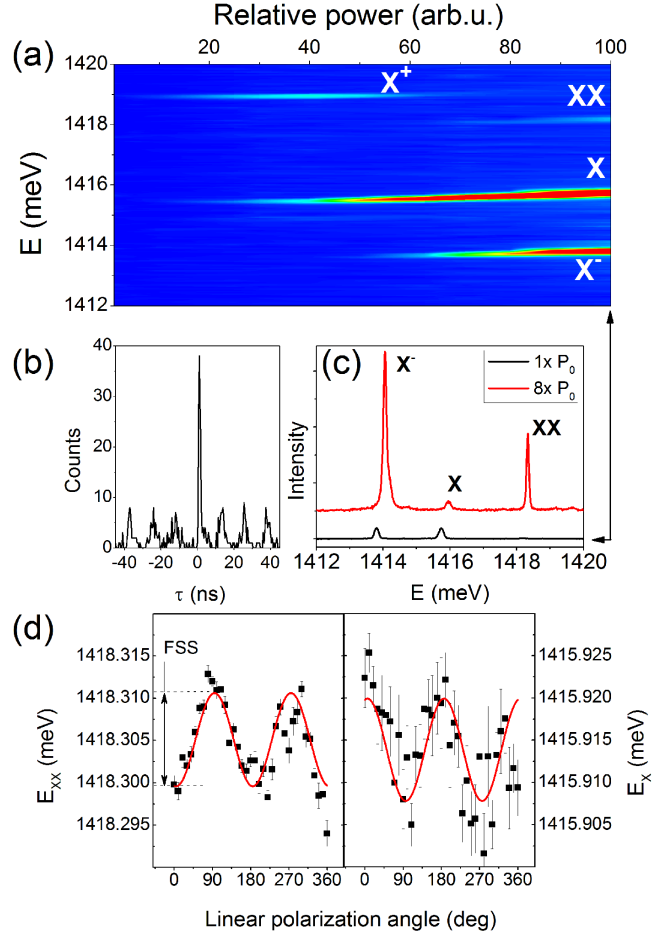

**FIG. 3.** Representative QD, see text. (a) Color map of excitation dependent photoluminescence at low power excitation conditions (b) Second-order correlation function between XX and X. (c) The bottom (black) spectrum is taken at excitation conditions used for the cross-correlation measurement shown in (b). The top (red) spectrum is taken at higher excitation power. The intensity level of X- and XX transitions is comparable to the level of transitions used for entanglement tests with QDs without dominant negative charging. (d) The fine-structure splitting of 11 µeV measured in XX and X recombinations confirming the type of transitions in the specific case.

### 2. Tuning QD charging

We stress that negative charging does not necessarily mean at all that a single QD is damaged or useless. The fine-structure splitting measurements showed that many of these dots are symmetric, as no clear FSS was resolved from the only clearly visible biexciton transition. In the further discussion we show that a different QD population methodology allows overcoming charging related issues nearly completely, and allows increasing significantly the number of good emitters.

Indeed, one of the efficient ways of modifying the QD charge state in a controlled manner is based on a dual-wavelength excitation[56]. The method takes advantage of deep levels present in the bandgap of GaAs. According to the excitation scenario discussed at the end of the previous section, excitation photons with energy higher than the acceptor-conduction band edge energetic separation create electron-hole pairs in a way that holes tend to remain trapped in acceptor levels, while electrons can be freed. By introducing a second excitation source with photons of the energy in-between the forbidden gap (in our case ~1180 meV), transitions from the valence band to the deep GaAs levels can take place. This excitation creates an excess concentration of holes which can eventually populate the QD and neutralise, or even, positively charge it. Thus ideally, the secondary excitation emission can be used as a sensitive QD charge tuning knob which does not affect (or has a minimal effect on) other QD properties, such as the fine-structure and the fine-structure splitting.

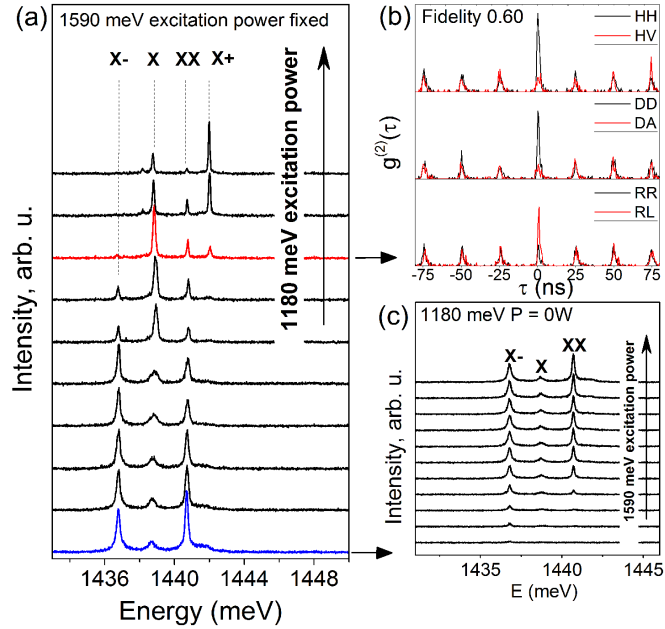

**FIG. 4.** (a) The set of spectra taken from a representative QD with variable dual wavelength excitation conditions. The above-bandgap excitation (1590 meV) was constant and only the 1180 meV excitation was increased. The increasing power of 1180 meV excitation is shown as a QD charge tuning mechanism. (b) Polarization-resolved second-order correlation curves measured from a QD with nearly neutral charge configuration. Polarization-entanglement is attested by the measured fidelity value of 0.600±0.025. (c) A single wavelength excitation power dependent spectra showing that the negative charge configuration is dominant at all conditions and no correlations between X and XX can be efficiently measured.

To test/demonstrate this charge tuning mechanism, a different, highly symmetric QD sample, the same reported in Ref. 14, was selected. FIG. 4 (a) shows spectra of a representative dot excited by a fixed above-bandgap excitation of 1590 meV and variable 1180 meV excitation. The bottom (blue) spectrum is taken when the secondary excitation was switched off. It is the typical spectrum observed from the majority of QDs. FIG. 4 (c) shows 1590 meV excitation power dependence where the neutral exciton intensity remains very small in the whole range, meaning that biexciton-exciton correlations cannot be analysed efficiently. When the secondary 1180 meV excitation is switched on, the charging of a QD changes gradually so that at a certain intensity the negative exciton is completely suppressed and the QD becomes nearly neutral with exciton and biexciton transitions dominant (the red curve). Further increase of 1180 meV pump charges the QD positively and a positive trion becomes dominant. At this stage the linewidth of the transitions became significantly narrower (50 μeV for the X+) comparing to a few hundred μeV width when a QD was negatively charged. This is probably the consequence of the neutralized electric field in the vicinity of a QD and a reduced spectral wandering[58]. By tuning a QD to a nearly neutral configuration, entanglement tests were carried out. FIG. 4 (b) shows the second-order correlation curves taken in linear, diagonal and circular bases. The calculated fidelity of the expected maximally entangled state was found to be 0.600±0.025 for this specific dot.

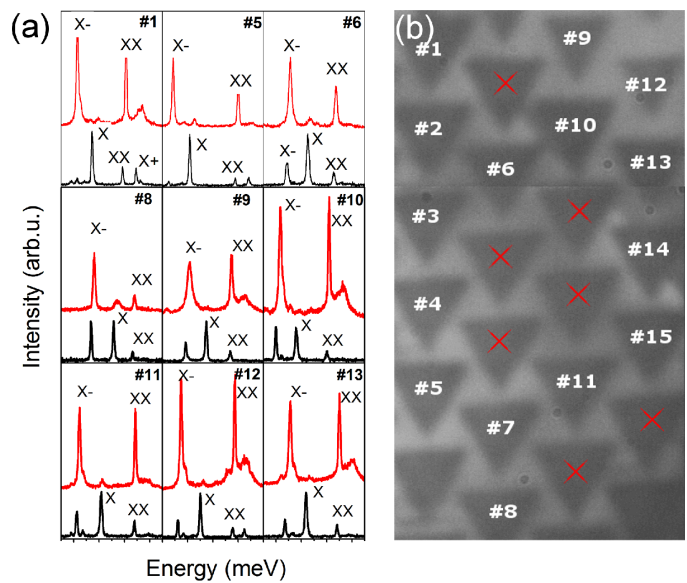

**FIG. 5.** (a) Nine representative spectra of QDs taken from the sample area shown in (b). The top (red) spectra are taken when QDs are excited only by a single wavelength excitation (1590 meV). The bottom (black) curves show the same QD spectra when the second excitation wavelength (1180 meV) is switch on. (b) The optical microscope image of the measured QDs. Red crosses indicate QDs that probably were damaged and no PL signal was obtained. The other QDs had dominant negative charge excitonic configurations.

The broad effectiveness of the charge tuning mechanism was tested by measuring the field of QDs presented in FIG.5 (b) where none of the 22 studied dots had a positive charge configuration (we did not perform entanglement measurement on this QDs field, as it would have required a time scale incompatible with the scope of this work). The crosses indicate QDs which did not have well pronounced excitonic features due to processing damage or (maybe) intrinsic defects, while the other shown and numbered dots had a negative charge configuration.

The corresponding spectra of nine representative QDs are presented by red (top) curves in FIG. 5 (a). The black (bottom) curves show the spectra when the secondary 1180 meV excitation is switched on. The power of both excitation sources was kept the same. In the studied region, the secondary excitation had a strong effect on most of the QDs by neutralising them to a level which could be successfully used for correlation measurements (even if we did not perform the full entanglement analysis). Considering the already high density of entangled photon emitters reported from this sample (see Ref. 14), this simple tuning method will be likely to enable much higher availability of potentially good entangled photon sources on this specific sample, as well as in other similar ones.

We should underline that, while this tuning method appeared to be highly efficient in some regions (and on different samples), in particular regions it had an insignificant effect possibly due to higher negative charging or different states in the band-gap involved, meaning that other quality improvements (e.g. sample processing, growth steps) are required. Our observations suggest that poor optical properties (negative charging, relatively broad linewidth) mainly arise due to configurations in the vicinity of a QD. A different QD population method, such as two-photon resonant excitation[59], would then allow overcoming these issues.

## III. CONCLUSIONS AN SUMMARY

We showed that entangled photon emission can be detected from site-controlled $In_{0.25}Ga_{0.75}As$ QDs designed in different ways, such as different QD thickness or shape. Different QD design allowed a coarse tuning of entangled photon emission in an overall range of ~80 meV. We conclude that the surfactant effects used in entangled photon emitter fabrication reported in Ref. 14 are helpful, however, not necessary. We presented two main factors that are currently limiting the purity of the polarization-entangled state and practical application of the Pyramidal QDs. First, a small, however, non-vanishing FSS was found in most of the dots. Second, a strong negative charging of non-resonantly excited QDs was shown as the main, however, not fundamental limitation of the current QD system. An efficient solution of this problem was demonstrated by the use of dual wavelength excitation, potentially improving the effectiveness of obtaining a high density of good entangled photon emitters on chip.

## ACKNOWLEDGEMENTS

This research was partly enabled by the Irish Higher Education Authority Program for Research in Third Level Institutions (2007-2011) via the INSPIRE programme, by Science Foundation Ireland under grants 10/IN.1/I3000 and 08/RFP/MTR/1659, and EU FP7 under the Marie Curie Reintegration Grant PERG07-GA-2010-268300. We thank K. Thomas for his support with the MOVPE system and V. Savona for useful discussions and inputs.

## REFERENCES

[1] P. G. Kwiat, K. Mattle, H. Weinfurter, A. Zeilinger, A. V. Sergienko and Y Shih, Phys. Rev. Lett. **75**, 4337 (1995).

[2] R. Horn, P. Abolghasem, B. J. Bijlani, D. Kang, A. S. Helmy and G. Wei, Phys. Rev. Lett. **108**, 153605 (2012).

[3] A. Orieux, A. Eckstein, A. Lemaître, P. Filloux, I. Favero1, G. Leo, T. Coudreau, A. Keller, P. Milman, and S. Ducci,Phys. Rev. Lett. **110**, 160502 (2013).

[4] R. M. Stevenson, R. J. Young, P. Atkinson, K. Cooper, D. A. Ritchie, and A. J. Shields, Nature **439**, 179 (2006);

[5] N. Akopian, N. H. Lindner, E. Poem, Y. Berlatzky, J. Avron, D. Gershoni, B. D. Gerardot, and P. M. Petroff, Phys. Rev. Lett. **96**, 130501 (2006);

[6] R. Hafenbrak, S. M. Ulrich, P. Michler, L. Wang, A. Rastelli, and O. G. Schmidt, New J. Phys. **9**, 315 (2007).

[7] A. Dousse, J. Suffczynski, A. Beveratos, O. Krebs, A. Lemaitre, I. Sagnes, J. Bloch, P. Voisin, and P. Senellart, Nature **466** , 217 (2010).

[8] M. A. M. Versteegh, M. E. Reimer, K. D. Jöns, D. Dalacu, P. J. Poole, A. Gulinatti, A. Giudice, and V. Zwiller, Nat. Commun. **5**, 5298, (2014).

[9] T. Huber, A. Predojević, M. Khoshnegar, D. Dalacu, P. J. Poole, H. Majedi, and G. Weihs, Nano Letters **14**, 7107 (2014).

[10] R. M. Stevenson, C. L. Salter, J. Nilsson, A. J. Bennett, M. B. Ward, I. Farrer, D. A. Ritchie, and A. J. Shields, Phys. Rev. Lett. **108**, 040503 (2012).

[11] R. M. Stevenson, J. Nilsson, A. J. Bennett, J. Skiba-Szymanska, I. Farrer, D. A. Ritchie, and A. J. Shields, Nature Commun **4**, 2859 (2013).

[12] S. J. Freedman and J. F. Clauser, Phys. Rev. Lett. **28**, 938-941 (1972).

[13] O. Benson, C. Santori, M. Pelton, and Y. Yamamoto, Phys. Rev. Lett. **84**, 2513-2516 (2000).

[14] G. Juska, V. Dimastrodonato, L. O. Mereni, A. Gocalinska, and E. Pelucchi, Nature Phot. **7**, 527 (2013).

[15] E. Pelucchi, V. Dimastrodonato, A. Rudra, K. Leifer, E. Kapon, L. Bethke, P. Zestanakis, and D. Vvedensky, Phys. Rev. B **83**, 205409 (2011).

[16] V. Dimastrodonato, E. Pelucchi, and D. D. Vvedensky, Phys. Rev. Lett. **108**, 256102 (2012)

[17] V. Dimastrodonato, E. Pelucchi, P. A. Zestanakis, and D. D. Vvedensky, Phys. Rev. B **87**, 205422 (2013).

[18] K. F. Karlsson, M. A. Dupertuis, D.Y. Oberli, E. Pelucchi, A. Rudra, P. O. Holtz, E. Kapon, Phys. Rev. B **81**, 161307(R) (2010).

[19] M. A. Dupertuis K. F. Karlsson, D. Y. Oberli, E. Pelucchi, A. Rudra, P.O. Holtz, and E. Kapon, Phys. Rev. Lett. **10**7, 127403 (2011).

[20] R. Singh, and G. Bester, Phys. Rev. Lett. **103**, 063601 (2009).

[21] T. Kuroda, T. Mano, N. Ha, H. Nakajima, H. Kumano, B. Urbaszek, M. Jo, M. Abbarchi, Y. Sakuma, K. Sakoda, I. Suemune, X. Marie, and T. Amand, Phys. Rev. B **88**, 041306 (2013).

[22] G. Juska, V. Dimastrodonato, L. O. Mereni, A. Gocalinska and E. Pelucchi, Nanoscale Res. Lett. **6**, 567 (2011).

[23] L. Auvray, H. Dumont, J. Dazord, Y. Monteil, J. Bouix, and C. Bru-Chevalier, J. Cryst. Growth **221**, 475 (2000).

[24] A. Aspect, J. Dalibard, P. Grangier and G. Roger, Opt. Commun. **49**, 429 (1984).

[25] T. Wilk, S. C. Webster, A. Kuhn and G. Rempe, Science **317**, 488 (2007).

[26] D. A. Kleinman, Phys. Rev. B **28**, 871 (1983).

[27] C. Ciuti, V. Savona, C. Piermarocchi, A. Quattropani, and P. Schwendimann, Phys. Rev. B **58**, 7926 (1998).

[28] C. Riva, F. M. Peeters, and K. Varga, Phys. Rev. B **61**, 13873 (1999).

[29] E. Tokunaga, A. L. Ivanov, S. V. Nair, and Y. Masumoto, Phys. Rev. B **59**, R7837 (1999).

[30] T. Ogawa and S. Okumura, Phys. Rev. B **65**, 035105 (2001).

[31] A. V. Filinov, C. Riva, F. M. Peeters, Yu. E. Lozovik, and M. Bonitz, Phys. Rev. B **70**, 035323 (2004).

[32] C. Schindler and R. Zimmermann, Phys. Rev. B **78**, 045313 (2008).

[33] Y. Benny, Y. Kodriano, E. Poem, S. Khatsevitch, D. Gershoni and P. M. Petroff, Phys. Rev. B **84**, 075473 (2011).

[34] R. M. Stevenson, A. J. Hudson, A. J. Bennett, R. J. Young, C. A. Nicoll, D. A. Ritchie, and A. J. Shields, Phys. Rev. Lett. **101**, 170501 (2008).

[35] D. F. V. James, P. G. Kwiat, W. J. Munro, and A. G. White, Physical Review A **64** (5), 052312 (2001).

[36] T. M. Stace, G. J. Milburn and C. H. W. Barnes, Phys Rev B **67**, 085317 (2003).

[37] M.O. Scully, B. Englert and H. Walter, Nature **351**, 6322 (1991).

[38] L.O. Mereni, V. Dimastrodonato, R.J. Young and E. Pelucchi, Appl. Phys. Lett. **94**, 223121 (2009).

[39] We used here a recently developed protocol, which includes gold bonding of our sample thermo-compressively, showing a dramatic improvement over the past recipes, as repetitive steps of cleanroom wax/epoxy for bonding are avoided. The new procedure results in better sample resilience by cancelling defect producing during the thermal cycling in the photoluminescence optical cryostat, generation of interface cracks and the like, all shortening the sample lifetime.

[40] J. B. Altepeter, E. R. Jeffrey, and P. G. Kwiat, Photonic State Tomography, Adv. At. Mol. Opt. Phy. **52**, 105 (2005).

[41] A. J. Hudson, R. M. Stevenson, A. J. Bennett, R. J. Young, C. A. Nicoll, P. Atkinson, K. Cooper, D. A. Ritchie, and A. J. Shields, Phys. Rev. Lett. **99**, 266802 (2007).

[42] A. Schliwa, M. Winkelnkemper, A. Lochmann, E. Stock, and D. Bimberg, Phys. Rev. B **80**, 161307 (2009).

[43] D. Y. Oberli, M. Byszewski, B. Chalupar, E. Pelucchi, A. Rudra, and E. Kapon, Phys. Rev. B **80**, 165312 (2009).

[44] X. Liu, N. Ha, H. Nakajima, T. Mano, T. Kuroda, B. Urbaszek, H. Kumano, I. Suemune, Y. Sakuma, and K. Sakoda, Phys. Rev. B **90**, 081301 (2014).

[45] C. D. Yerino, P. J. Simmonds, B. Liang, D. Jung, C.Schneider, S. Unsleber, M. Vo, D. L. Huffaker, S. Höfling, M. Kamp, and M. L. Lee, Appl. Phys. Lett. **105,** 251901 (2014).

[46] V. Mlinar and A. Zunger, Phys. Rev. B **79**, 115416 (2009).

[47] G. Juska, V. Dimastrodonato, L. O. Mereni, T. H. Chung, A. Gocalinska and E. Pelucchi, B. Van Hattem, P. Corfdir, and M. Ediger, Phys. Rev. B **89**, 205430 (2014)

[48] R. Trotta, E. Zallo, C. Ortix, P. Atkinson, J. D. Plumhof, J. van den Brink, A. Rastelli, and O. G. Schmidt, Phys. Rev. Lett. **109**, 147401 (2012).

[49] M. Baier, F. Findeis, A. Zrenner, M. Bichler, and G. Abstreiter, Phys. Rev. B **64**, 195326 (2001).

[50] A. Schliwa, M. Winkelnkemper, and D. Bimberg, Phys. Rev. B **79**, 075443 (2009).

[51] E. Poem, J. Shemesh, I. Marderfeld, D. Galushko, N. Akopian, D. Gershoni, B. D. Gerardot, A. Badolato, and P. M. Petroff, Phys. Rev. B **76**, 235304 (2007).

[52] T. Warming, E. Siebert, A. Schliwa, E. Stock, R. Zimmermann, and D. Bimberg, Phys. Rev. B **79**, 125316 (2009).

[53] Y. Igarashi, M. Shirane, Y. Ota, M. Nomura, N. Kumagai, S. Ohkouchi, A. Kirihara, S. Ishida, S. Iwamoto, S. Yorozu, and Y. Arakawa, Phys. Rev. B **81**, 245304 (2010).

[54] T. Kuroda, T. Belhadj, M. Abbarchi, C. Mastrandrea, M. Gurioli, T. Mano, N. Ikeda, Y. Sugimoto, K. Asakawa, N. Koguchi, K. Sakoda, B. Urbaszek, T. Amand, and X. Marie, Phys. Rev. B **79**, 035330 (2009).

[55] B. Urbaszek, R. J. Warburton, K. Karrai, B. D. Gerardot, P. M. Petroff, and J. M. Garcia, Phys. Rev. Lett. **90**, 247403 (2003).

[56] E. S. Moskalenko, V. Donchev, K. F. Karlsson, P. O. Holtz, B. Monemar, W. V. Schoenfeld, J. M. Garcia, and P. M. Petroff, Phys. Rev. B **68**, 155317 (2003).

[57] M. K. Hudait, P. Modak, S. Hardikar, and S. B. Krupanidhi, J. Appl. Phys. **83**, 4454 (1998).

[58] A. V. Kuhlmann, J. Houel, A. Ludwig, L. Greuter, D. Reuter, A. D. Wieck, M. Poggio, and R. J. Warburton, Nature Phys. **9**, 570 (2013).

[59] M. Muller, S. Bounouar, K. D. Jons, M. Glassl, and P. Michler, Nature Phot. **8**, 224 (2014).